\documentclass[preprint,
 amsmath,amssymb,
 aps,
 pra,
 floatfix
]{revtex4-2}

\usepackage{graphicx}
\usepackage{dcolumn}
\usepackage{bm}
\usepackage[english]{babel}
\makeatletter
\@namedef{l@en}{\l@english}
\makeatother
\usepackage{hyperref}
\usepackage{pgfplots}

\usepackage{booktabs}
\usepackage{algpseudocode}

\pgfplotsset{compat=1.18}

\begin{document}

\preprint{APS/123-QED}

\title{Efficient Quantum Error Correction from Three Dimensional Qubit Control}

\author{Kevin Yipu Wu}
\email{ypwk@uw.edu}
\author{Ohik Kwon}
\author{Maxwell F. Parsons}
\affiliation{ECE Department, University of Washington}

\date{\today}

\begin{abstract}
High-rate quantum low-density parity-check (qLDPC) codes can substantially reduce qubit overhead relative to surface codes, but their advantage depends on efficiently realizing nonlocal syndrome extraction. We study the \([[144,12,12]]\) bivariate bicycle code on a neutral-atom architecture with native three-dimensional (3D) geometry, comparing planar and 3D embeddings while holding the code fixed. We characterize spatial efficiency using the logical-qubit density, defined as the number of encoded logical qubits per unit spatial footprint. Because the optical controller's field of view limits the transverse extent of an array, this metric estimates the number of logical qubits that can be accommodated within a fixed optical field of view. The 3D embedding achieves approximately \(4\times\) greater areal logical-qubit density than the planar layout and \(42\times\) greater than a surface-code baseline. It also reduces the bivariate bicycle syndrome-extraction time by roughly \(2\times\) compared to a planar baseline, with fewer movement operations and substantially shorter atom-transport distance. Native 3D geometry can improve both the packing density and executable realization of nonlocal qLDPC codes, making practical performance depend jointly on code structure, optical geometry, transport scheduling, and hardware-level noise. This motivates further development of control techniques in 3D. 
\end{abstract}
\maketitle

\section{Introduction}
Fault-tolerant quantum computation aims to execute long algorithms reliably despite noise. Physical qubits are inherently error-prone: they decohere over time, and operations introduce additional errors. Quantum error correction addresses this problem by encoding logical qubits across many physical qubits and repeatedly extracting error syndromes during computation. The resulting overhead in qubit count, control complexity, and execution time remains one of the primary obstacles to scalable quantum computing.

This overhead depends both on the physical hardware and on the structure of the error-correcting code. These factors are closely coupled. A code that is efficient in theory may lose much of its advantage if its stabilizer measurements are difficult to realize on the underlying hardware.

Encoding rate is defined as the number of logical qubits encoded per physical qubit. High-rate quantum low-density parity-check (qLDPC) codes can substantially reduce qubit overhead relative to surface codes \cite{bravyiHighthresholdLowoverheadFaulttolerant2024, panteleevDegenerateQuantumLDPC2021, panteleevAsymptoticallyGoodQuantum2022, leverrierSmallQuantumTanner2025}. However, many high-rate codes require nonlocal stabilizer measurements between qubits that are widely separated in space \cite{baspinConnectivityConstrainsQuantum2022}. On hardware platforms with constrained connectivity, implementing these interactions can introduce substantial routing, scheduling, and control overhead.

This motivates codesigning quantum codes and hardware architectures. Neutral-atom quantum processors suit this approach because qubits can be dynamically arranged and transported within programmable optical tweezer arrays \cite{everedHighfidelityParallelEntangling2023, pauseSuperchargedTwodimensionalTweezer2024, chungFaulttolerantOperationMaterials2025, linAIEnabledParallelAssembly2025, pichardRearrangementIndividualAtoms2024, norciaIterativeAssembly$^171$$mathrmYb$2024}. Entangling interactions can then be selectively activated through Rydberg coupling. Existing neutral-atom systems, face scaling constraints associated with optical power and trap depth \cite{manetschTweezerArray61002025, holmanTrappingSingleAtoms2026}. Each optical tweezer must provide sufficient intensity to confine an atom reliably, so increasing the number of independently generated traps increases the total laser-power requirement, making scaling costly and technically difficult. Extending these architectures into three-dimensional (3D) geometries may increase atom count without a commensurate increase in laser power, by engineerring diffraction propagation to allow reuse of laser power across multiple planes \cite{barredoSyntheticThreedimensionalAtomic2018}. 

We study the implementation of a high-rate qLDPC code, the $[[144,12,12]]$ bivariate bicycle (BB) code \cite{bravyiHighthresholdLowoverheadFaulttolerant2024}, or the ``gross code", on a neutral-atom architecture with native 3D geometry. Prior work has studied syndrome-extraction schedules for this BB code under several architectural assumptions \cite{viszlaiQSIEVEEfficientQLDPC2026, pooleArchitectureFastImplementation2025}, specifically in two-dimensional (2D) atom-array geometries. Since syndrome extraction is repeated throughout fault-tolerant computation and contributes substantially to quantum-error-correction runtime, its implementation cost is a central determinant of practical overhead. To our knowledge, no prior work has studied 3D embeddings for this code. Architecturally, our approach combines features of static and dynamically reconfigurable neutral-atom implementations. Rather than relying entirely on long-range interactions or shuttling individual qubits between dedicated zones, we keep the data registers stationary and collectively translate the check registers between interaction configurations, exploiting the finite Rydberg blockade radius to execute multiple interactions at each stop.

We compare planar and 3D embeddings of the same BB code under a common hardware and noise model. By holding the code fixed while varying the embedding geometry and hardware-level schedule, we isolate the effect of geometry on routing overhead, syndrome-extraction depth, and simulated logical performance. The 3D embedding reduces transport overhead and syndrome-extraction time by bringing atoms into closer proximity while maintaining comparable simulated logical error-rate curves. In addition, as the field of view of both objective lenses and spatial light modulators are limited, there is a limit to the number of qubits and logical qubits we may pack into a single field of view. We evaluate areal and volumetric logical qubit density, which measures the number of logical qubits that can be supported in a unit area (\textmu m\(^2\)) or unit volume (\textmu m\(^3\)), respectively. Utilizing a 3D geometry allows us to pack more logical qubits into the same transverse footprint. These results suggest that native 3D geometry and control can reduce the implementation overhead of nonlocal qLDPC syndrome extraction, but that logical performance depends jointly on code structure, hardware geometry, and scheduling.

\begin{figure}
    \centering
    \includegraphics[width=\linewidth]{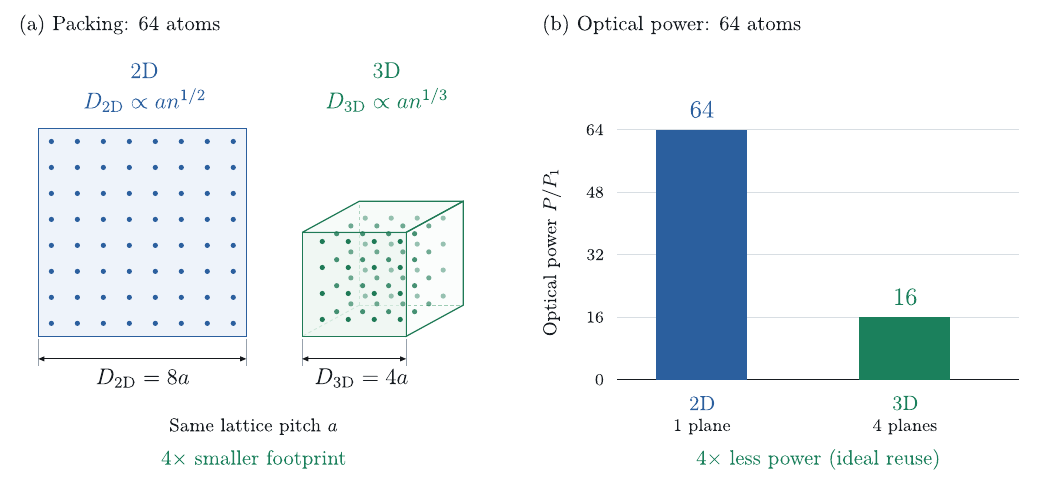}
    \caption{Packing and ideal optical-power benefits of 3D atom arrays.
    (a) At equal lattice pitch $a$, 64 atoms occupy transverse footprints
    of $(8a)^2$ in 2D and $(4a)^2$ in 3D.
    (b) At equal trap depth and waist, ideal reuse of trapping light across
    four planes reduces the required power from $64P_1$ to $16P_1$,
    where $P_1$ is the incident power per independent trap.
    This fourfold reduction assumes engineered optical reuse;
    optical losses and trap nonuniformity limit practical gains.}
    \label{fig:fig_dimensional_scaling}
\end{figure}

\section{Background}

Neutral-atom quantum processors provide a useful setting to study this tradeoff because their connectivity is reconfigurable through optical tweezer motion and Rydberg interactions. They encode qubits in the internal electronic states of individual atoms confined by optical tweezer traps. These traps are formed by tightly focused laser beams, allowing arrays of atoms to be assembled with programmable spatial configurations. In typical implementations, qubits are stored in long-lived hyperfine ground states, while entangling interactions are mediated by excitation to high-lying Rydberg states. When one atom is excited to a Rydberg state, it shifts the energy levels of nearby atoms, suppressing simultaneous excitation within a characteristic blockade radius. This Rydberg blockade mechanism enables fast, distance-limited two-qubit gates.

In this platform, the array geometry can be dynamically reconfigured. Arrays are generated using spatial light modulators, and atoms can be rearranged or transported using acousto-optic deflectors \cite{picardThreedimensionalAcoustoopticDeflector2026}, enabling selective addressing and programmable connectivity. 
Unlike solid-state platforms with fixed layouts, interactions can be induced by bringing atoms into proximity or by activating Rydberg coupling within a finite range. However, this nonlocal connectivity comes at a steep cost — shuttling operations are orders of magnitude slower than gate operations in state of the art experimental neutral atom platforms \cite{bluvsteinArchitecturalMechanismsUniversal2026}. This motivates reducing total runtime by optimizing atom movement.

Prior work has studied syndrome-extraction schedules for this BB code under several architectural assumptions \cite{viszlaiQSIEVEEfficientQLDPC2026, pooleArchitectureFastImplementation2025}, specifically in two-dimensional (2D) atom-array geometries. These approaches span static implementations that exploit long-range Rydberg interactions and dynamic implementations that use atom transport to realize nonlocal checks. Here, we consider an intermediate architecture in which the data registers remain stationary while the check registers are transported collectively between interaction configurations. At each configuration, the finite Rydberg interaction range allows multiple compatible check--data interactions to be executed without additional transport. We study how embedding this architecture natively in three dimensions changes the cost of syndrome extraction.

\section{Methods}
We study two planar and one three-dimensional (3D) embedding of the $[[144,12,12]]$ bivariate bicycle (BB) code, the gross code, together with twelve distance-11 surface codes. We compare the planar and 3D implementations of the BB code under a shared hardware and noise model so that changes in performance arise from geometry and syndrome-extraction scheduling instead of differences in code parameters or decoding assumptions. The other planar implementation and the surface codes are included as baselines. We choose twelve distance-11 surface codes, as it encodes the same number of logical qubits at a similar distance to the BB code. 

\subsection{Hardware Assumptions}
\label{sec:hardware_assumptions}

The simulations are parameterized using near-term timing estimates for a $^{87}$Rb neutral-atom processor (Table \ref{tab:timing_model}). The 3D architecture is a near-term architectural extrapolation rather than a model of an existing device. Code to reproduce the simulations in this paper is available on Zenodo to adjust assumptions, see \cite{kevinyipuwuYpwkBivariatebicyclecode3DV12026}. In particular, the model assumes:
\begin{enumerate}
    \item programmable generation of high uniformity three-dimensional optical tweezer arrays,
    \item spatially addressable one- and two-qubit operations throughout a 3D volume,
    \item collision-free atom transport in all three spatial dimensions with parallel rectilinear motion, and
    \item Single-shot multi-layer measurements within a multi-plane geometry. 
\end{enumerate}

The first three of these capabilities have partial experimental precedent in existing neutral-atom systems \cite{barredoSyntheticThreedimensionalAtomic2018, holmanTrappingSingleAtoms2026, manetschTweezerArray61002025, zhangScaledLocalGate2024, picardThreedimensionalAcoustoopticDeflector2026}, although no current platform simultaneously realizes the full architecture assumed here. This model evaluates how how native 3D geometry and control affects the implementation cost of nonlocal qLDPC syndrome extraction, and whether there exist reasons beyond simple qubit count scaling to further develop 3D control capabilities.

Transport operations are constrained to rectilinear motion patterns consistent with acousto-optic tweezer control \cite{everedHighfidelityParallelEntangling2023}. Parallel movement is therefore limited by shared transport directions, making the spatial organization of stabilizer interactions an important component of syndrome-extraction scheduling. Although our syndrome extraction scheme does not use axial transport, axial motion is required to assemble defect-free 3D arrays.

\begin{table}
    \centering
    \begin{tabular}{c|c}
        \textbf{Parameter} & \textbf{Value / Model} \\
        \hline
        Readout duration & $160\,\mu s$ \cite{sheaSubmillisecondNondestructiveTimeresolved2020} \\
        Single-qubit gate duration & $5\,\mu s$ \cite{bluvsteinQuantumProcessorBased2022} \\
        Two-qubit Rydberg gate duration & $0.27\,\mu s$ \cite{everedHighfidelityParallelEntangling2023} \\
        Maximum transport acceleration & $v_{accel} = 0.02 \,\mu m/\mu s^2$ \cite{xuConstantoverheadFaulttolerantQuantum2024} \\
        Tweezer transfer overhead & $T_{\mathrm{transfer}} = 50\,\mu s$ \cite{bluvsteinQuantumProcessorBased2022} \\
        Rydberg blockade radius & $R_b = 8.02\,\mu\mathrm{m}$\\ 
        \(T_1\) and \(T_2\) time & $4\,s\quad\mathrm{and}\quad1.5\,s$  \cite{bluvsteinQuantumProcessorBased2022} 
    \end{tabular}
    \caption{Default timing and interaction parameters used in the dynamic-device model.}
    \label{tab:timing_model}
\end{table}

Table \ref{tab:timing_model} contains default timing and interaction parameters used in the dynamic device model. We model transport time for a jerk-free cubic movement profile as in \cite{viszlaiQSIEVEEfficientQLDPC2026} and  \cite{xuConstantoverheadFaulttolerantQuantum2024}, where for a displacement of \(\Delta x\) and \(\Delta y\) in two dimensions, the time required is 
\begin{equation}
    T_{\mathrm{transport}}(d) = \sqrt{\frac{6\Delta x}{v_{\mathrm{accel}}}} + \sqrt{\frac{6\Delta y}{v_{\mathrm{accel}}}}.
\end{equation}
A Rydberg blockade radius of $8.02 \mu$m was computed for the $50s_{1/2}$ level in $^{87}$Rb using the following approximation to the long-range interaction energy: 
\begin{equation}
    V(R) = \frac{C_6}{R^6}
\end{equation}
for the non-resonant regime. We sum over dipole-coupled intermediate pair states using the Alkali-Rydberg Calculator Python package \cite{sibalicARCOpensourceLibrary2017} to compute the van der Waals interaction coefficient $C_6$, then compute the blockade radius by equating the interaction shift to the excitation linewidth:
\begin{equation}
    R_b = \Big(\frac{|C_6|}{\hbar \Gamma}\Big)^{1/6}.
\end{equation}

Of the architectural assumptions listed above, simultaneous multi-layer measurement represents the largest departure from demonstrated capability. 
Tweezer-array generation, three-dimensional atom transport, and spatially addressed gates in 3D geometries each have at least partial experimental precedent
\cite{barredoSyntheticThreedimensionalAtomic2018,holmanTrappingSingleAtoms2026,manetschTweezerArray61002025,zhangScaledLocalGate2024,picardThreedimensionalAcoustoopticDeflector2026}.
In contrast, simultaneous fluorescence readout of multiple axial planes with sufficient suppression of inter-plane crosstalk has not yet, to our knowledge, been demonstrated for neutral-atom arrays. If simultaneous multi-layer readout is unavailable, a possible relaxation is to measure the check-qubit layers sequentially, with the imaging system refocused between axial planes. Recently, Ref.~\cite{zengFastNondestructiveReadout2026} showed that high fidelity nondestructive readout can be performed in 10s of microseconds using a multimode fiber array attached to avalanche photodiodes, making this refocusing approach less temporally expensive.  


\subsection{Embedding and Scheduling the Gross Code}

\begin{figure}
    \centering
    \includegraphics[width=\linewidth]{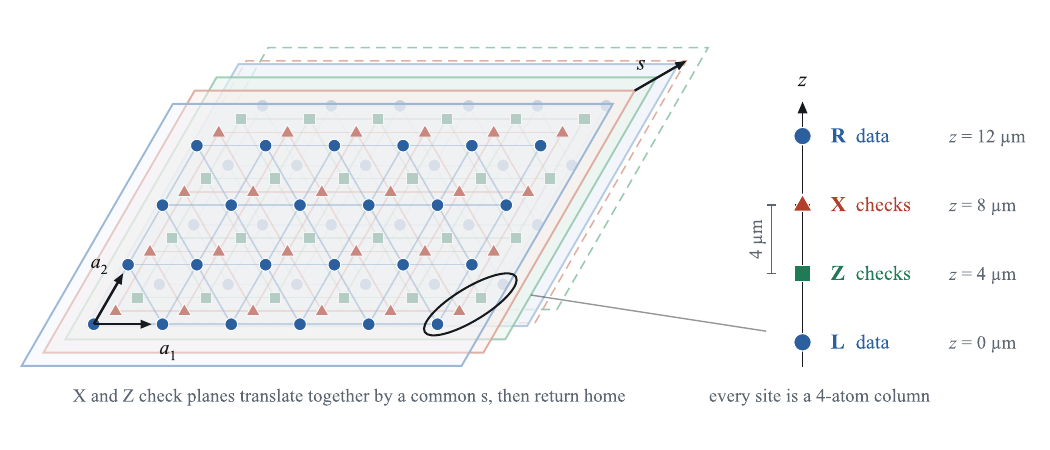}
    \caption{Schematic of triangular 3D layout with four layers leading to efficient implementation of the BB syndrome extraction circuit. In our layout, \(|a_1| = |a_2| = 4\) \textmu m. In this figure, 6 \(\times\) 4 of each 12 \(\times\) 6 register shown. Dashed outlines mark the check planes at each stop. The four planes coincide in x-y; they are drawn with a small parallax offset so each remains visible.}
    \label{fig:triangular_plan_overview}
\end{figure}

Parallel atom transport in acousto-optic tweezer systems is naturally constrained to shared rectilinear motion patterns. The embedding problem therefore consists of arranging stabilizer interactions such that many checks  can be executed simultaneously with minimal transport overhead, while respecting the native dynamic trap array geometry produced by acousto-optic deflectors. 

We consider three types of layouts for the gross code: two planar layouts in 2D and a layout in 3D. Because the same BB code is used in all layouts, differences in cycle time, transport overhead, and logical performance arise from the embedding geometry and resulting schedules. 
The ordering of CNOTs in the syndrome extraction circuits will change the effective distance of the circuit \cite{strikisHighperformanceSyndromeExtraction2026}. 
Therefore, two different syndrome extraction circuits for the same code could have very different threshold behavior. 
To isolate the impact of platform geometry from the impact of circuit distance in our logical error rate simulations, we choose a non-interleaved distance-10 left-right (LR) circuit using the framework of Ref.~\cite{strikisHighperformanceSyndromeExtraction2026} and schedule it in both 2D and 3D.
We refer to the 2D layout and scheduler as ``\textit{LR planar}", and the 3D layout and scheduler as ``\textit{LR triangular}". 
In addition, we include a 2D layout and scheduler from existing work \cite{viszlaiQSIEVEEfficientQLDPC2026}, which we refer to as the ``\textit{systolic}" circuit. 

Finally, as the speed of scheduling and density performance of the different layouts depend heavily on the choice of spacing, we maintain comparable lattice spacing between both planar and 3D layouts to make a fair comparison. We make sure in both embeddings that the distance between all pairs of qubits are at least \(4\ \mathrm{\mu m}\). In all cases in this study, we impose a geometric exclusion constraint between simultaneously executed two-qubit gates during scheduling, to avoid crosstalk during entangling gates.

\subsubsection{Systolic Baseline Layout and Scheduler}

The systolic baseline follows a two-dimensional embedding of the gross code together with a dynamic systolic scheduler proposed in Ref.~\cite{viszlaiQSIEVEEfficientQLDPC2026}. Qubits are arranged to reduce the number of unique transport vectors required during syndrome extraction. This produces an arrangement in a square planar grid, with each class of data and check qubit, L or R and X or Z respectively, having different offsets to avoid collisions during transport. Due to the embedding geometry, if the locations of each unique transport vector are visited in sequence during syndrome extraction, then all of the long range interactions are at some point made local during the tour. Therefore, the traveling salesperson algorithm is applied to solve this problem, with qubit travel time as the minimization objective. See Ref.~\cite{viszlaiQSIEVEEfficientQLDPC2026} for a more in-depth description of this layout. 

\subsubsection{LR Planar Baseline Layout and Scheduler}

The LR planar implementation places the four 72-atom registers of the \([[144,12,12]]\) BB code---left data, \(X\) checks, \(Z\) checks, and right data---side by side in a single plane.  The atoms form a \(48\times6\) rectangular array with \(4\,\mu\mathrm{m}\) spacing, occupying a coordinate span of \(188\,\mu\mathrm{m}\times20\,\mu\mathrm{m}\). A single-AOD scheduler picks up both check registers once and carries them through an exact 24-stop tour.  The tour minimizes transport time while preserving the optimized CNOT order on every qubit; stops on disjoint qubits may be interleaved, even across the nominal \(X/Z\) phase boundary.  At each
stop, the applicable CNOTs are executed in parallel, after which the checks return home and are measured together.  The resulting check-only
syndrome-round schedule is repeated for the 12 rounds of the logical-memory experiment. See Appendix \ref{apdx:bb-lr-planar} for more information about this layout and scheduler. 

\subsubsection{LR Triangular 3D Layout and Scheduler}

In the LR triangular layout, the \(L\), \(Z\), \(X\), and \(R\) registers occupy four 4 \textmu m-spaced planes, each containing the same 4 \textmu m-pitch triangular lattice. During syndrome extraction, the X and Z-check registers are loaded once into mobile tweezers and translated together in the $xy$-plane using a single  transport waveform. Each interaction group can be executed over a finite region of check-register displacements for which all intended check–data pairs lie within the Rydberg interaction range. When these valid regions overlap, multiple compatible interaction groups can be executed at the same transport stop, reducing the number of distinct displacements required. We use a precedence-constrained search to identify a 19-stop schedule while preserving the prescribed gate order on every qubit. Further details of the layout and scheduler are given in Appendix~\ref{apdx:bb-lr-triangular}; Figure~\ref{fig:triangular_plan_overview} illustrates the geometry and transport scheme.

\subsection{Simulation Details}
\label{sec:simulation_details}

Performance is evaluated using a quantum-memory experiment. Twelve logical $\lvert 0\rangle_L$ states are initialized and evolved under repeated rounds of syndrome extraction. After $d$ rounds, where $d$ is the full code distance, the logical state is measured and compared with the expected logical outcome over a sweep of physical noise strengths. In this case, $d=11$ for the surface code and $d=12$ for the bivariate bicycle code.

For each layout, we generate an explicit hardware-level schedule specifying atom transport, gate execution, idling, reset, and measurement. 
Surface-code schedules are produced using a greedy local scheduler, while the BB schedules are constructed as described above. 
The resulting schedules are compiled into Stim-compatible noisy circuits \cite{gidneyStimFastStabilizer2021} and sampled using Sinter.
Surface-code decoding is performed using minimum-weight perfect matching through PyMatching
\cite{higgottSparseBlossomCorrecting2025}, while the BB codes are decoded using belief propagation with ordered-statistics post-processing (BP-OSD) \cite{panteleevDegenerateQuantumLDPC2021}.

We use a circuit-level stochastic Pauli noise model controlled by a single dimensionless circuit-noise scaling parameter $p$. One- and two-qubit gates are followed by one- and two-qubit depolarizing channels, respectively, each with total nonidentity Pauli probability $p$. 
Reset and measurement faults are modeled as Pauli errors with probability $p$. 
These discrete operation-level channels are interpreted as effective error models for the complete physical operation, including decoherence and other imperfections incurred during the
execution of the corresponding gate, reset, or measurement. 
We therefore do not apply an additional duration-dependent coherence channel during these operations.

Schedule-dependent storage decoherence is accumulated only during intervals in which quantum information is retained without an active gate, reset, or measurement operation. In particular, this includes idle periods and atom transport. For an elapsed storage interval of duration $t$, the combined effects of $T_1$ relaxation and $T_2$ dephasing are computed via the Pauli twirling approximation \cite{gellerEfficientErrorModels2013}:
\begin{align}
    p_X(t) = p_Y(t)
        &= \frac{1-e^{-t/T_1}}{4}, \\
    p_Z(t)
        &= \frac{1-2e^{-t/T_2}+e^{-t/T_1}}{4}.
\end{align}
Thus, an interval of duration $t$ is represented by an independent single-qubit Pauli channel with probabilities $p_X(t)$, $p_Y(t)$, and $p_Z(t)$. 
This construction avoids double counting decoherence already included phenomenologically in the discrete operation-error channels.

The coherence times are scaled with the global noise parameter according to
\begin{equation}
    T_i(p)
    =
    \frac{p_{\mathrm{ref}}}{p}T_{i,\mathrm{ref}},
\end{equation}
with
\begin{equation}
    p_{\mathrm{ref}} = 6.25\times10^{-3}, \qquad
    T_{1,\mathrm{ref}} = 4~\mathrm{s}, \qquad
    T_{2,\mathrm{ref}} = 1.5~\mathrm{s}.
\end{equation}
The reference values are chosen to provide conservative neutral-atom operating scales \cite{bluvsteinQuantumProcessorBased2022, manetschTweezerArray61002025}. For $p=0$, all stochastic operation, coherence, and transfer errors are disabled.

Atom transport is treated as a storage interval of finite duration. 
Consequently, a transported qubit accumulates the same duration-dependent $T_1/T_2$ decoherence as an idle qubit over the corresponding elapsed time. 
We do not include an additional movement-specific depolarizing channel in the default model.
Instead, each pickup or drop-off between static and mobile tweezers is assigned an independent transfer fault with probability
\begin{equation}
    p_{\mathrm{TRF}} = 0.16p,
\end{equation}
corresponding to a transfer-fault probability of $10^{-3}$ per transfer at $p=p_{\mathrm{ref}}$, a quantity seen experimentally \cite{manetschTweezerArray61002025}. The transfer fault is modeled as a single-qubit depolarizing channel with total nonidentity probability $p_{\mathrm{TRF}}$. 
Coherence accumulated during finite pickup, drop-off, routing, and other transport-related delays is accounted for through the corresponding scheduled storage intervals. 
Coherent neutral-atom transport has been demonstrated experimentally using dynamical-decoupling techniques \cite{bluvsteinQuantumProcessorBased2022}.

The parameter $p$ is a dimensionless global noise-scaling parameter and should not be interpreted as a device-specific gate-error probability. 
We scale discrete operation-error probabilities linearly with $p$ and coherence rates $1/T_i$ linearly with $p$, such that all modeled error mechanisms vanish together as $p\to0$. 
This convention defines a one-parameter family of circuit-level noise models for comparing schedules and layouts. It is not intended to predict how gate fidelity, coherence times, and transfer fidelity co-vary in a particular experimental system.

Under this convention, differences in scheduled idle and transport duration produce additional storage-decoherence penalties, whereas the durations of active gates, resets, and measurements do not incur a separate $T_1/T_2$ channel because their total error is already represented by the corresponding
discrete operation-level fault channel. 
The resulting model therefore separates effective operation faults from schedule-induced storage decoherence.

The complete model combines stochastic operation faults, Pauli-twirled storage decoherence during idle and transport intervals, and tweezer-transfer faults. 
It does not explicitly model leakage, atom loss or erasure information, crosstalk, coherent control errors, or correlated faults, and should not be interpreted as a  detailed device-level noise model of a particular neutral-atom processor.

The raw logical failure probability over an \(r=d\)-round memory experiment is estimated as
\begin{equation}
    \widehat{p}_{\mathrm{fail}}
    =
    \frac{E}{N},
\end{equation}
where \(N\) is the number of sampled shots and \(E\) is the number containing at least one logical-observable error. We report the corresponding per-round logical error rate,
\begin{equation}
    \widehat{p}_{\mathrm{LER}}
    =
    1-\left(1-\widehat{p}_{\mathrm{fail}}\right)^{1/r}.
\end{equation}

For each value of \(p\), Monte Carlo sampling is continued until
\(N_{\mathrm{err}}=400\) logical failures have been observed. The resulting relative statistical uncertainty is therefore approximately
\begin{equation}
    \frac{\sigma_{\widehat{p}_{\mathrm{LER}}}}
         {\widehat{p}_{\mathrm{LER}}}
    \sim
    \frac{1}{\sqrt{N_{\mathrm{err}}}}
    \approx 5\%,
\end{equation}
so points across the physical-error-rate sweep have comparable relative sampling uncertainty.

The BB codes are decoded using BP-OSD as implemented in
\texttt{stimbposd}, with min-sum belief propagation, \(10{,}000\) BP
iterations, combination-sweep OSD, and OSD order \(10\). The surface-code
results use MWPM through PyMatching
\cite{higgottSparseBlossomCorrecting2025}.

\section{Results and Discussion}

\begin{figure}[ht]
\centering
\includegraphics[width=\textwidth]{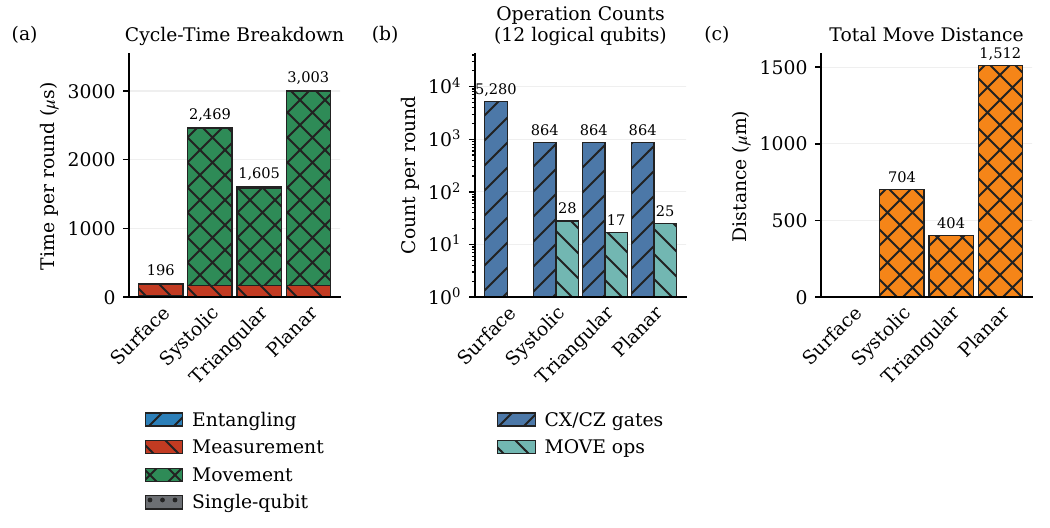}
\caption{Syndrome-extraction schedule comparison for twelve planar
distance-11 surface-code patches and three implementations of the
\([[144,12,12]]\) bivariate bicycle (BB) code: LR-triangular, LR-planar,
and systolic.
(a) Cycle-time decomposition for one syndrome-extraction round.
The surface-code baseline is fastest because its stabilizer interactions are
geometrically local and require no atom transport. Among the BB
implementations, the LR-triangular layout has the shortest cycle time,
primarily because of its reduced routing overhead.
(b) Operation counts. All BB implementations require the same
number of entangling gates, \(864\), but differ in the number of movement
operations required by their physical schedules. The surface code requires no movement since all entangling operations are local. 
(c) Total atom-transport distance. The LR-triangular layout
requires the shortest total movement distance, \(404~\mu\mathrm{m}\),
compared with \(1512~\mu\mathrm{m}\) for LR-planar and
\(704~\mu\mathrm{m}\) for the systolic implementation, illustrating that
the cycle-time reduction arises from the physical realization of the
nonlocal interactions rather than a reduction in entangling-gate count.}
\label{fig:scheduling_comparison}
\end{figure}

We evaluate each layout along three axes: spatial overhead (Fig.~\ref{fig:footprint_template}), temporal overhead (Fig.~\ref{fig:scheduling_comparison}), and logical performance (Fig.~\ref{fig:logical_error_rate}). 

\subsection{Spatial Overhead}
\label{sec:spatial_overhead}

\begin{figure*}[htbp]
\centering
\includegraphics[width=0.5\linewidth]{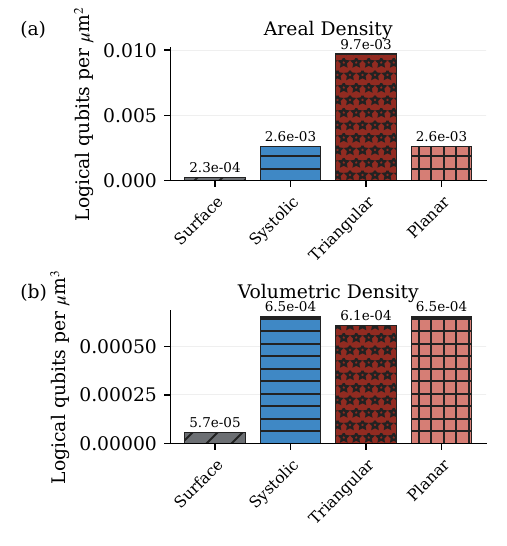}
\caption{Spatial-footprint comparison for twelve logical qubits encoded
using a planar distance-11 surface-code baseline and three
\([[144,12,12]]\) bivariate bicycle (BB) implementations:
LR-triangular, LR-planar, and systolic.
Projected area and corresponding areal logical-qubit density are shown
alongside the three-dimensional bounding-box volume and volumetric density.
The LR-triangular layout has the smallest projected footprint and the highest
areal density, reducing the transverse area by approximately a factor of four
relative to the planar BB implementations. When the axial extent is included,
the volumetric advantage is smaller: under the assumed \(4~\mu\mathrm{m}\)
thickness for planar layouts, the density of the LR-triangular implementation is about the same as the density of the LR-planar layout. The comparison illustrates
the trade-off introduced by the 3D embedding, which uses additional axial extent to reduce the required transverse footprint.}
\label{fig:footprint_template}
\end{figure*}

Spatial overhead is evaluated using two complementary metrics: the projected
planar footprint and the three-dimensional bounding-box volume. The projected
footprint measures the area occupied in the transverse plane and is relevant
when optical field of view, planar addressing, or imaging area limits the
available hardware. The volumetric metric additionally accounts for the
axial extent of the layout.

As shown in Fig.~\ref{fig:footprint_template}, the LR-triangular layout has
the smallest projected footprint, occupying \(1237~\mu\mathrm{m}^2\),
compared with \(4608~\mu\mathrm{m}^2\) for LR-planar,
\(4608~\mu\mathrm{m}^2\) for the systolic implementation, and
\(52630~\mu\mathrm{m}^2\) for the surface-code baseline. Since the BB
implementations encode the same number of logical qubits, this corresponds
to an areal logical-qubit density of approximately
\(9.7\times10^{-3}\) logical qubits per \(\mu\mathrm{m}^2\) for
LR-triangular, compared with \(2.6\times10^{-3}\) for LR-planar and
\(2.6\times10^{-3}\) for the systolic layout. The LR-triangular geometry
therefore provides approximately a \(4\times\) increase in areal density
relative to the planar BB implementations and a substantially larger
increase relative to the surface-code baseline.

The advantage disappears when the full
three-dimensional extent is included. The LR-triangular, LR-planar, and systolic BB layouts occupy bounding-box volumes of approximately
\(19785~\mu\mathrm{m}^3\), \(18432~\mu\mathrm{m}^3\), and
\(18432~\mu\mathrm{m}^3\), respectively, corresponding to volumetric
logical-qubit densities of approximately \(6.1\times10^{-4}\),
\(6.5\times10^{-4}\), and \(6.5\times10^{-4}\) logical qubits per
\(\mu\mathrm{m}^3\). For the planar layouts, we assign an axial extent of
\(4~\mu\mathrm{m}\) when converting projected area to bounding-box volume, and include a \(4~\mu\mathrm{m}\) margin around the extent of the LR-triangular layout in each dimension.  
Under this convention, the LR-triangular layout is approximately the same density as the LR-planar and systolic layout in volume. 

The projected-area comparison more directly captures the benefit of the
triangular embedding for architectures in which transverse optical access
and imaging area are the dominant spatial constraints. A purely volumetric
comparison can obscure this advantage because it treats axial and transverse
extent equivalently, whereas extending an array along the axial direction
can introduce additional addressing and measurement constraints that are
not represented by bounding-box volume alone.

\subsection{Temporal Overhead}
\label{sec:temporal_overhead}

Temporal overhead is evaluated from the duration of one syndrome-extraction
round and its operation-level decomposition. The surface-code baseline has
the shortest cycle time, \(196~\mu\mathrm{s}\), and requires no atom
transport. The BB implementations are slower because their nonlocal
stabilizer interactions require additional routing and movement.

Among the BB implementations, the LR-triangular layout has the shortest
syndrome-extraction time, \(1605~\mu\mathrm{s}\), compared with
\(3002~\mu\mathrm{s}\) for LR-planar and \(2469~\mu\mathrm{s}\) for the
systolic implementation. Thus, the triangular layout reduces the cycle time
by approximately a factor of two relative to LR-planar and is also faster
than the systolic schedule.

All three BB implementations perform the same number of entangling
operations, \(864\), so the differences in cycle time arise primarily from
their physical routing schedules rather than from differences in the number
of stabilizer interactions. The LR-triangular implementation requires
\(17\) movement operations with a total transport distance of
\(404~\mu\mathrm{m}\), compared with \(25\) movements and
\(1512~\mu\mathrm{m}\) for LR-planar. The systolic implementation requires
\(28\) movements totaling \(704~\mu\mathrm{m}\). The triangular layout
therefore substantially reduces the transport distance relative to the
planar LR implementation, while also requiring fewer movement operations.

These results show that the temporal overhead of the BB syndrome-extraction
circuit is strongly influenced by the physical realization of its nonlocal
interactions. Although the BB implementations require the same number of
entangling operations, their routing costs differ substantially, with the
LR-triangular geometry providing the lowest cycle time among the BB layouts
considered here.

In this work, we assume that decoding can be completed within the
millisecond-scale timescale of syndrome extraction. This requirement is more
demanding for the BB code than for the surface-code baseline. BP-OSD includes
an ordered-statistics post-processing stage involving Gaussian elimination,
which has been identified as a potential obstacle to real-time decoding at
neutral-atom or trapped-ion cycle times of order \(1~\mathrm{ms}\)
\cite{roffeBiastailoredQuantumLDPC2023}. Recent work has investigated faster
decoding approaches, including beam-search decoders
\cite{yeBeamSearchDecoder2025}, hardware-oriented implementations of BP-OSD
\cite{mullerImprovedBeliefPropagation2025}, and parallelized BP variants that
avoid Gaussian elimination \cite{wangFullyParallelizedBP2026}. Our logical
error-rate results use BP-OSD with a fixed offline decoding budget and
therefore do not include decoder latency in the syndrome-extraction cycle
time. Real-time implementation would additionally require a decoder capable
of keeping pace with the corresponding syndrome-extraction schedule.

\subsection{Interaction Ordering and Physical-Layout Trade-offs}
\label{sec:ordering-layout-tradeoffs}

\begin{figure}
    \centering
    \includegraphics[width=\linewidth]{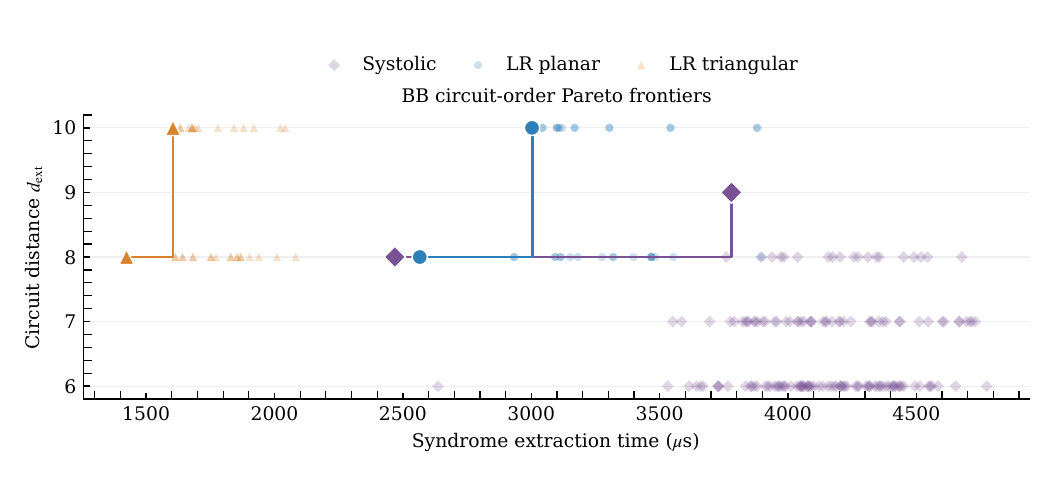}
    \caption{Trade-off between syndrome-extraction time and extended-code
      distance \(d_{\mathrm{ext}}\) for systolic, LR-planar, and
      LR-triangular implementations.
      Each faint marker represents a distinct valid ordering of the
      stabilizer interactions, while the enlarged markers and solid lines
      identify the non-dominated points within each implementation family.
      The points corresponding to the LR-planar and LR-triangular implementations use the same
      syndrome-extraction circuit and span the same 36 interaction orderings; only the
      physical arrangement and resulting transport schedule differ between the two sets.
      Extended-code distances are estimated using BP-OSD. Note that changing the interaction ordering could change the extended distance of the code, as well as the syndrome extraction time. 
    }
    \label{fig:circuit-order-pareto}
\end{figure}

Figure~\ref{fig:circuit-order-pareto} separates two effects of the interaction ordering: its influence on hook-error propagation and the physical cost of realizing the resulting schedule. 
Following Ref.~\cite{strikisHighperformanceSyndromeExtraction2026}, we characterize the former using the extended-code distance \(d_{\mathrm{ext}}\). 
For a given interaction ordering, a fault on a stabilizer auxiliary qubit can propagate through subsequent entangling gates and produce a residual error on multiple data qubits. 
Because the propagation path depends on the sequence of stabilizer interactions, different interaction orderings can produce different sets of residual errors. 
The systolic and LR-planar implementations both exhibit this trade-off: different valid interaction orderings produce different residual-error sets and hence different values of \(d_{\mathrm{ext}}\), while also requiring different physical schedules.

The extended-code construction incorporates these residual errors into a code-capacity distance calculation by introducing an additional variable for each residual error. 
When the complete residual-error set \(\mathcal{E}\) is included, the resulting quantity \(d_{\mathrm{ext}}(\mathcal{E})\) is the full extended-code distance.
For the non-interleaved syndrome-extraction circuits considered here, Ref.~\cite{strikisHighperformanceSyndromeExtraction2026} shows that
\begin{equation}
    d_{\mathrm{circ}} = d_{\mathrm{ext}}(\mathcal{E}),
\end{equation}
where \(d_{\mathrm{circ}}\) is the circuit distance. Thus, within this circuit model, \(d_{\mathrm{ext}}\) provides a computationally convenient characterization of how the interaction ordering affects circuit-level fault tolerance. 

For the comparisons in the remainder of this study, we select operating points from the resulting time--distance trade-off. 
For the systolic approach, we choose to use the fastest $d_{\mathrm{ext}} = 8$ circuit, corresponding to the circuit generated by the approach of Ref.~\cite{viszlaiQSIEVEEfficientQLDPC2026}, which serves as our baseline. 
For the LR-planar and LR-triangular implementations, we use the fastest ordering with \(d_{\mathrm{ext}}=10\) for each physical layout.

The LR-triangular implementation uses the same LR syndrome-extraction circuit and the same interaction orderings as the LR-planar implementation. 
Consequently, each corresponding ordering has the same residual-error set and the same \(d_{\mathrm{ext}}\). 

Overall, the interaction ordering determines the residual-error set and therefore the attainable circuit distance, while the physical layout
determines how efficiently that ordering can be executed. 
The family-specific Pareto frontiers provide a basis for selecting an operating point that balances syndrome-extraction time and circuit distance.

\subsection{Logical Protection}

\begin{figure}[thbp]
\centering
\includegraphics[width=0.5\columnwidth]{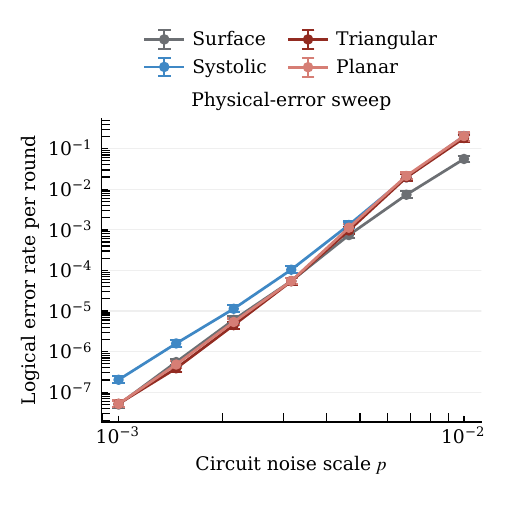}

\caption{
Logical error rate per round of syndrome extraction for the studied layouts under the circuit-level noise model. A shot is counted as a logical failure if at least one of the twelve tracked logical observables is decoded incorrectly. We first estimate the experiment-level probability \(p_{\mathrm{any}}\) of one or more logical failures over \(R\) syndrome extraction rounds, then report the equivalent per-round rate \(p_{\mathrm{any,round}} = 1 - (1-p_{\mathrm{any}})^{1/R}\). We include 95\(\%\) Wilson confidence intervals. 
}
\label{fig:logical_error_rate}
\end{figure}

See Fig.~\ref{fig:logical_error_rate} for a plot of the logical error rate of our Monte Carlo simulation. We find that as the circuit noise scale \(p\) decreases, the error rate corresponding to dephasing and decoherence from idling and movement decrease, leading the two LR circuits to converge to the performance of the surface code baseline. As predicted by the Pareto frontier plot (Fig.~\ref{fig:circuit-order-pareto}), the systolic approach has lower extended circuit distance, leading to worse logical protection overall versus the two LR circuits, despite faster scheduling when compared to the LR planar schedule.  

\section{Conclusion}

We have shown that native three-dimensional geometry provides a concrete architectural advantage for implementing nonlocal quantum LDPC codes on neutral-atom hardware. By codesigning the embedding and syndrome-extraction schedule of the \([[144,12,12]]\) bivariate bicycle code, we find that the 3D layout substantially reduces both spatial and temporal resource requirements relative to a planar implementation of the same code. Because the two BB layouts have identical code parameters, these gains arise from differences in the embedding geometry and the resulting hardware-level schedule.

These results point toward broader advantages of three-dimensional qubit control. Spatial dimension directly constrains the achievable rate--distance tradeoff for geometrically local codes. For a CSS code equivalent to one that is geometrically local in \(D\) dimensions with bounded stabilizer weight \(\Delta\), generalized Bravyi--Poulin--Terhal-type bounds give \(d \lesssim n^{1-1/D}\) \cite{postemaExistenceCharacterizationBivariate2026}. This exponent changes substantially with \(D\): \(1/2\) at \(D=2\), the familiar surface-code scaling; \(2/3\) at \(D=3\); and \(3/4\) at \(D=4\). The \([[144,12,12]]\) code studied here is itself an instance of this tradeoff. Because its stabilizers have weight six and due to its algebraic properties, it is equivalent to a code that is geometrically local in four spatial dimensions \cite{postemaExistenceCharacterizationBivariate2026}. From this perspective, the routing overhead quantified in Sec.~IV~B can be viewed as the cost of realizing this higher-dimensional connectivity on lower-dimensional hardware. Our results show concretely, for this code and syndrome-extraction protocol, that supplying an additional physical dimension can reduce that cost.

A further motivation for three-dimensional control is access to codes that have no two-dimensional counterpart. Passive protection of quantum information is perhaps the sharpest case. A recent construction of a 3D Pauli stabilizer Hamiltonian whose ground space encodes a qubit with exponentially long memory lifetime when coupled to a bath at nonzero temperature \cite{balasubramanianPassiveSelfcorrectingQuantum2026} therefore provides additional motivation for studying hardware that can realize genuinely three-dimensional interaction geometries. Whether such constructions can be implemented efficiently on neutral-atom hardware, and what architectural advantages 3D would provide in that setting, remain open questions. Yet another motivatin for 3D control concerns non-Clifford resource production. Magic-state distillation based on the \([[15,1,3]]\) quantum Reed--Muller code exploits transversal non-Clifford structure to produce high-fidelity \(T\) resource states \cite{bravyiUniversalQuantumComputation2005,campbellMagicStateDistillationAll2012}. Color-code and subsystem color-code constructions provide related approaches in which three-dimensional gauge-fixing procedures expose universal transversal gates while using lower-weight gauge measurements \cite{bombinGaugeColorCodes2015,brownFaulttolerantErrorCorrection2016}. These constructions suggest another setting in which access to native three-dimensional connectivity could be useful. However, determining whether 3D neutral-atom hardware reduces the implementation overhead of such protocols, and by how much relative to optimized 2D realizations, requires a separate architecture-level analysis analogous to the one carried out here for BB syndrome extraction.

Several directions therefore remain for extending this work. It would be valuable to study asymptotically good high-rate qLDPC families with longer codes, such as lifted-product codes \cite{panteleevDegenerateQuantumLDPC2021}, in a 3D setting to determine whether the scheduling advantages observed here generalize to other algebraic code families. A simple scaling argument suggests that the scheduling advantage of 3D could grow slowly with code length, as \(n^{1/12}\): 
\begin{align*}
    &L_{2D} \sim n^{1/2}, \quad L_{3D} \sim n^{1/3} \\ 
    \Rightarrow \quad  &T_{2D} / T_{3D} \approx \sqrt{L_{2D} / L_{3D}} \sim n^{1/4-1/6} = n^{1/12}. 
\end{align*}

However, establishing this behavior will require explicit embeddings and schedules for larger instances. Integrating such layout-level gains into full-stack resource estimates for fault-tolerant workloads, including applications such as Shor's algorithm \cite{cainShorsAlgorithmPossible2026}, would provide a more complete assessment of their system-level significance. Similar architecture-aware studies of self-correcting memories, gauge-fixing protocols, and resource-state factories could clarify whether the advantages found here extend beyond quantum-memory syndrome extraction.

Taken together, our results establish a concrete advantage of 3D qubit control: for the BB code studied here, the additional spatial degree of freedom reduces the routing required to realize the same Tanner graph, shortens the syndrome-extraction round, and correspondingly reduces the movement- and idle-induced noise accumulated during each round. Thus, 3D hardware can lower the physical overhead of executing at least some forms of nonlocal qLDPC syndrome extraction without changing the underlying code. More generally, treating hardware geometry and code structure as coupled design choices when evaluating fault-tolerant architectures, while determining how broadly the advantages of native 3D extend remains an important direction for future work.

\section*{Acknowledgements}

The authors would like to thank Chinmay Nirkhe and Michael Beverland for helpful discussions. K.Y.W. and O.K. were partially supported through the Advancing Quantum-Enabled Technologies (AQET) traineeship program at the University of Washington, National Science Foundation Award DGE-2021540.

\section*{Data Availability}

The data that support the findings of this article are openly available \cite{kevinyipuwuYpwkBivariatebicyclecode3DV12026}.

\bibliography{references}

\appendix

\section{Scheduling}

We distinguish an \emph{interaction visit} from a physical \textsc{Move} operation. 
An interaction visit is a displacement at which one or more entangling gates are executed.  
A \textsc{Move} operation is a nonzero transport leg between two consecutive displacements, including the initial departure from home and the final return home.

Let $\bm{s}_1,\ldots,\bm{s}_{N_{\mathrm{vis}}}$ denote the interaction visits, and define
\[
    \bm{s}_0=\bm{s}_{N_{\mathrm{vis}}+1}=\bm{0},
\]
where $\bm{0}$ is the home displacement.  The total transport time is
\[
    T_{\mathrm{move}}
    =
    \sum_{k=0}^{N_{\mathrm{vis}}}
    \tau\!\left(
        \left\lVert
            \bm{s}_{k+1}-\bm{s}_k
        \right\rVert
    \right).
\]
A physical \textsc{Move} is emitted only when $\bm{s}_{k+1}\neq\bm{s}_k$.

\subsection{LR Planar layout and syndrome extraction schedule}
\label{apdx:bb-lr-planar}

This configuration implements the \([[144,12,12]]\) bivariate bicycle (BB) code using the left--right (LR) circuit construction.

The code is defined on a \(12\times 6\) periodic
index set by
\[
  A(x,y)=x^3+y+y^2,
  \qquad
  B(x,y)=y^3+x+x^2.
\]

Its atoms are divided into four registers: the left and right data registers,
denoted \(L\) and \(R\), and the \(X\)- and \(Z\)-check-ancilla registers,
denoted \(X\) and \(Z\). Each register contains \(12\times6=72\) atoms.
Consequently, the layout contains 144 data atoms and 144 check ancillas, for
a total of 288 atoms.

Writing a local register index as
\[
  q=6i+j,
  \qquad
  0\leq i<12,\quad 0\leq j<6,
\]
the position of the atom at site \((i,j)\) in register
\(s\in\{L,X,Z,R\}\) is
\[
  \mathbf{r}_{s}(i,j)
  =
  \bigl(16i+\delta_s,\;4j,\;0\bigr)\,\mu\mathrm{m},
\]
where
\[
  \delta_L=0,\qquad
  \delta_X=4\,\mu\mathrm{m},\qquad
  \delta_Z=8\,\mu\mathrm{m},\qquad
  \delta_R=12\,\mu\mathrm{m}.
\]

Thus, each logical lattice site is expanded into four adjacent register
positions ordered \(L\)-\(X\)-\(Z\)-\(R\) along the \(x\) direction. The
factor-of-four stretching of the logical \(i\) coordinate leaves a
\(4\,\mu\mathrm{m}\) separation between the \(R\) register at site \(i\) and
the \(L\) register at site \(i+1\). The complete arrangement is therefore a
\(48\times6\) array with \(4\,\mu\mathrm{m}\) nearest-neighbor spacing. Its
coordinate span is
\[
  192\,\mu\mathrm{m}\times24\,\mu\mathrm{m},
\]
and hence the footprint convention used in the study assigns it an area of
\(4608\,\mu\mathrm{m}^2\). All physical coordinates have \(z=0\); the
\(4\,\mu\mathrm{m}\) thickness used by the footprint analysis is only an
effective thickness for reporting volumetric density.

For each check type, the six color ticks each decompose into two
rigid displacement groups.  The combined $X$--$Z$ scheduling problem
therefore contains
\[
    N_{\mathrm{vis}}^{\mathrm{planar}}
    = 2 \times 6 \times 2
    = 24
\]
interaction visits.

The ordering of these 24 visits is obtained by an exact
precedence-constrained Held--Karp dynamic program.  A dependency is
inserted whenever reversing two visits would reverse the prescribed
CNOT order on a shared check or data qubit.  Visits acting on
qubit-disjoint sets may be interleaved, including across the nominal
boundary between the $Z$- and $X$-check phases.

Writing the selected interaction displacements as
$\bm{s}_1,\ldots,\bm{s}_{24}$, with
$\bm{s}_0=\bm{s}_{25}=\bm{0}$, the routing objective is
\[
    T_{\mathrm{move}}^{\mathrm{planar}}
    =
    \sum_{k=0}^{24}
    \tau\!\left(
        \left\lVert
            \bm{s}_{k+1}-\bm{s}_k
        \right\rVert
    \right).
\]
The implemented route contains 24 interaction visits and 25 nonzero
transport legs: one move from home to the first visit, 23 moves
between consecutive visits, and one final move back home.  It
therefore contains
\[
    N_{\textsc{Move}}^{\mathrm{planar}}=25
\]
physical \textsc{Move} operations.

At each interaction visit, all CNOTs belonging to the corresponding
rigid displacement group are executed in parallel.  After the
twenty-fourth interaction visit, the check array is returned home by
the twenty-fifth and final \textsc{Move}.  All check atoms are then
transferred back into static traps, the final Hadamards are applied
to the $X$-check qubits, and the $X$- and $Z$-check qubits are
measured and reset together.

\subsection{LR Triangular Layout and Syndrome Extraction Schedule}
\label{apdx:bb-lr-triangular}

The LR triangular layout places each of the \(L\), \(Z\), \(X\), and \(R\) registers on a separate triangular-lattice plane. For site indices
\((i,j)\), the common in-plane coordinate is
\[
  \mathbf{r}_{\parallel}(i,j)
  =
  \left(4i+2j,\;2\sqrt{3}\,j\right)\,\mu\mathrm{m},
\]
corresponding to primitive lattice vectors of length \(4\,\mu\mathrm{m}\)
separated by \(60^\circ\). The four planes are uniformly separated by
\(4\,\mu\mathrm{m}\), with register heights
\[
  z_L=0,\qquad
  z_Z=4\,\mu\mathrm{m},\qquad
  z_X=8\,\mu\mathrm{m},\qquad
  z_R=12\,\mu\mathrm{m}.
\]
Thus, the register order is \(L\)-\(Z\)-\(X\)-\(R\) from bottom to top.
This ordering was selected by enumerating the assignments of the four registers to the four fixed heights and evaluating each feasible assignment with the same routing objective. 
The LR coloring and the prescribed qubit-local gate order are held fixed during this optimization. The term ``triangular'' refers to the lattice within each plane; the full layout is a four-plane, three-dimensional configuration.

The LR triangular scheduler uses a single common transport waveform for the two check registers. 
Both check registers are loaded at the beginning of a round and translated together in the \(xy\) plane, while their \(z\) coordinates remain fixed. 
The data registers remain stationary. 
Because both check registers receive the same displacement, distinct transport legs are serialized. At each interaction visit, the currently ready interactions are executed in one or more qubit-disjoint parallel batches, with no additional transport between batches.

The scheduler allows any interaction group to be executed at any displacement for which all of its
intended check--data pairs lie within the blockade range and the complete atom arrangement satisfies the minimum-spacing constraint. 
For a rigid interaction group \(g\), these feasible common displacements define an interaction region
\[
  \mathcal{D}_g
  =
  \left\{
      \mathbf{s}\in\mathbb{R}^{2}:
      d_{cd}(\mathbf{s})\leq r_{\mathrm{int}}
      \text{ for every }(c,d)\in g
  \right\},
\]
where \(r_{\mathrm{int}}\) is chosen slightly below the maximum interaction
range. Intersections of these regions allow multiple interaction groups to
be executed from a single interaction waypoint.

Routing is subject to the prescribed gate order on every check and data
qubit. Two groups may be interleaved only when doing so does not reverse the
order of two interactions incident on the same qubit. 
Subject to the prescribed gate order, the blockade-overlap
construction certifies that the interaction groups can be serviced
in a minimum of
\[
    N_{\mathrm{vis}}^{\mathrm{tri}}=16
\]
interaction visits.  These visits are selected from a finite witness
set consisting of interaction-region centers and pairwise boundary
intersections that satisfy the global minimum-spacing constraint.
The home displacement $\bm{0}$ is included among the candidate
witness points, but it is not selected as an interaction visit in
the implemented route.

A subsequent dynamic program chooses the minimum-time route over
this finite witness set.  Writing the selected interaction
displacements as $\bm{s}_1,\ldots,\bm{s}_{18}$, with
$\bm{s}_0=\bm{s}_{19}=\bm{0}$, the movement objective is
\[
    T_{\mathrm{move}}^{\mathrm{tri}}
    =
    \sum_{k=0}^{18}
    \tau\!\left(
        \left\lVert
            \bm{s}_{k+1}-\bm{s}_k
        \right\rVert
    \right).
\]
Because the selected route begins away from home, contains no
zero-length transitions between consecutive visits, and returns
home after its final visit, it contains
\[
    N_{\textsc{Move}}^{\mathrm{tri}}
    = N_{\mathrm{vis}}^{\mathrm{tri}}+1
    = 17
\]
physical \textsc{Move} operations.

At each interaction visit, every currently ready interaction covered
by that witness point is divided into qubit-disjoint gate batches.
These batches are executed without additional transport between
them.  After all 16 interaction visits have been serviced, the
nineteenth \textsc{Move} returns both check registers to their home
coordinates, after which they are transferred back to static traps
and read out together.

\end{document}